\documentclass[12pt]{article}

\usepackage{amssymb}
\usepackage{amsmath}
\usepackage{amscd}
\usepackage{mathrsfs}
\usepackage{amssymb}
\usepackage{feynmf}
\usepackage{graphicx}
\usepackage{soul}

\usepackage{tikz}

\def\dj{\hbox{d\kern-0.347em \vrule width 0.3em height 1.252ex depth
-1.21ex \kern 0.051em}}

\numberwithin{equation}{section}

\allowdisplaybreaks

\begin{document}

\setlength{\oddsidemargin}{0cm}
\setlength{\baselineskip}{7mm}

%%%%%%%%%TEXT%%%%%%%%%%%%%%%%%%%%%%%%%%%%%%%%%%%%%%%%%%%%%%%%%%%%%%%%%
%%%%%%%%%%%%%%%%%%%%%%%%%%%%%%%%%%%%%%%%%%%%%%%%%%%%%%%%%%%%%%%%%%%%%B
%%%%%%%%%%%%%%%%%%          title page       %%%%%%%%%%%%%%%%%%%%%%%%%
%%%%%%%%%%%%%%%%%%%%%%%%%%%%%%%%%%%%%%%%%%%%%%%%%%%%%%%%%%%%%%%%%%%%%%

\thispagestyle{empty}
\setcounter{page}{0}

\begin{flushright}
CERN-PH-TH/2013-162
\end{flushright}

\vspace*{1cm}

\begin{center}
{\bf \Large Color-Kinematics Duality in Multi-Regge Kinematics }

\vspace*{0.5cm}

{\bf \Large and Dimensional Reduction}

%\vspace*{0.4cm}

%{\bf \Large of Inelastic Amplitudes}

%\vspace*{0.3cm}
%
%{\bf \Large  Amplitudes}
%

\vspace*{0.3cm}

Henrik Johansson,$^{a,}$\footnote{\tt henrik.johansson@cern.ch}
Agust\'{\i}n Sabio Vera,$^{b,}$\footnote{\tt a.sabio.vera@gmail.com}
\\
Eduardo Serna Campillo,$^{c,}$\footnote{\tt eduardo.serna@usal.es} 
Miguel \' A. V\'azquez-Mozo$^{c, }$\footnote{\tt Miguel.Vazquez-Mozo@cern.ch}

\end{center}

\vspace*{0.0cm}

\begin{center}
$^{a}${\sl Theory Division, Physics Department, CERN, 
CH-1211 Geneva 23, Switzerland
}

$^{b}${\sl Instituto de F\'{\i}sica Te\'orica UAM/CSIC \&
Universidad Aut\'onoma de Madrid\\
C/ Nicol\'as Cabrera 15, E-28049 Madrid, Spain
}

$^{c}${\sl Departamento de F\'{\i}sica Fundamental \& IUFFyM,
 Universidad de Salamanca \\ 
 Plaza de la Merced s/n,
 E-37008 Salamanca, Spain
  }
\end{center}

\vspace*{0.5cm}

\centerline{\bf \large Abstract}

In this note we study the applicability of the color-kinematics duality to the scattering of two distinguishable 
scalar matter particles with gluon emission in QCD, or graviton emission in Einstein gravity. Previous analysis suggested that direct use of the Bern-Carrasco-Johansson double-copy prescription to matter amplitudes does not reproduce the gravitational amplitude in multi-Regge kinematics. 
This situation, however, can be avoided by extensions to the gauge theory, while maintaning the same Regge limit. 
Here we present two examples of these extensions: 
the introduction of a scalar contact interaction and the relaxation of the 
distinguishability of the scalars. In both cases new diagrams allow for a full reconstruction of the correct Regge limit on the gravitational side. Both modifications correspond to theories obtained by dimensional reduction from higher-dimensional gauge theories.

\vspace*{0.5cm}

\noindent

\newpage

\setcounter{footnote}{0}

\section{Introduction}

Deeper understanding of the relationship between gravity and gauge theories has been a longstanding problem in
high-energy theoretical physics, both in the strong-weak aspect of AdS/CFT~\cite{AdS/CFT} and in the weak-weak setting of perturbative calculations~\cite{review_gauge_gravity}. 
Early glaring evidence of a perturbative relationship came from string theory, where closed-string graviton
vertex operators can be seen as a direct product of open-string gauge-field vertex operators. This structure leads to 
a relation between gravity and gauge theory tree amplitudes known as the Kawai-Lewellen-Tye (KLT) relations~\cite{KLT}. In the 90's this problem was further studied~\cite{BernGravity} strengthening the idea that, at the level of scattering amplitudes, gravity is in some loose sense the ``square'' of a gauge theory. 

This idea was made precise following a proposal put forward by Bern, Carrasco, and one of the current authors (BCJ), who recognized that the underlying mechanism is a duality between color and kinematics present in gauge theory~\cite{BCJ}. The color-kinematics duality then generates gravity amplitudes by replacing the color factors in a gauge-theory amplitude with kinematic numerator functions depending on particle momenta and states, giving a double-copy representation of gravity amplitudes.
Much of the power of the proposal stems from the fact that it is expected to directly generalize to loop 
amplitudes~\cite{BCJLoop} (see also~\cite{BDHK}), opening the door to the riches of interesting perturbative calculations~\cite{LoopNumerators,RecentLoopNumerators,N>=4SG,N=4SG}.  

At tree level, the inclusion of general matter states and interactions in the color-kinematics duality is an open problem. In a recent paper~\cite{SVSCVM2}, the duality was studied in the context of inelastic amplitudes involving scalar particles in multi-Regge kinematics (the relation between multi-Regge kinematics~\cite{Bartels:2008ce,Bartels:2008sc} and supergravity amplitudes in the BCJ context has been explored in~\cite{Bartels:2012ra}). It was shown that an initial application of the BCJ 
color-kinematics duality to the scattering of two scalar particles with gluon emission in scalar QCD only
retrieves part of the gravitational amplitude.  More precisely, the part that was correctly obtained corresponds to the square of two Lipatov's QCD 
emission vertices~\cite{BFKL1,BFKL2,BFKL3}. The terms 
crucial for the cancellation of simultaneous divergences in overlapping channels~\cite{Lipatov:2011ab,Lipatov:1982vv,Lipatov:1982it,Lipatov:1991nf,Bartels:2012ra,SVSCVM}, as required by unitarity
(Steinmann relations~\cite{Steinmann}),  were absent in this application of the duality.
While the main line of studies of color-kinematics duality deals with pure (super)-Yang-Mills theories in various dimensions, where the double-copy prescription is proven~\cite{BDHK}, the calculation of Ref.~\cite{SVSCVM2} incorporated additional minimally-coupled matter states in Yang-Mills theory. As this is a setup that is outside of the standard application of color-kinematics duality, it is perhaps not unexpected that a refined double-copy prescription is needed.

In this note we revisit the problem pointed out in Ref.~\cite{SVSCVM2}, approaching it with two different modifications
of the theory. 
Firstly, we consider the scattering of two distinguishable scalars in Yang-Mills theory, where the scalars live in the adjoint representation. As a crucial new element, we introduce the quartic matter self-coupling characteristic of the
bosonic sector of $\mathcal{N}=2$ supersymmetric Yang-Mills theory. Secondly, we repeat the calculation of Ref.~\cite{SVSCVM2}, only this time considering {\em identical} adjoint scalars. 
In both cases, the color-kinematics duality reproduces, in the Regge limit, the gravitational amplitude as computed in \cite{SVSCVM}, including those terms responsible for the fulfillment of the Steinmann relations. 
The $D=4$ Yang-Mills + scalar theories studied here are via dimensional reduction directly related to pure Yang-Mills theory in $D=6$ and $D=5$ dimensions, respectively. Similarly, the theories can be thought of as originating from the bosonic sector of ${\cal N}=2$  super-Yang-Mills theory. This explains why the inclusion of matter in these cases is straightforward from the perspective of color-kinematics duality.

\section{Color-kinematics duality with scalar matter }

Our aim is to study gauge-theory scattering of two scalar particles with the emission of a gluon; and later on, gravity scattering of two scalars with the emission of a graviton. These have momenta $p_{1}$, $p_{2}$ (incoming scalars), $p_{3}$, $p_{4}$ (outgoing scalars), and $p_{5}$ (emitted gluon/graviton), which are all taken to enter the diagram.  Before particularizing to the case of distinguishable or indistinguishable scalars and matter self-interactions, we carry out a general analysis.

\begin{figure}
\centerline{\includegraphics[width=5.0in]{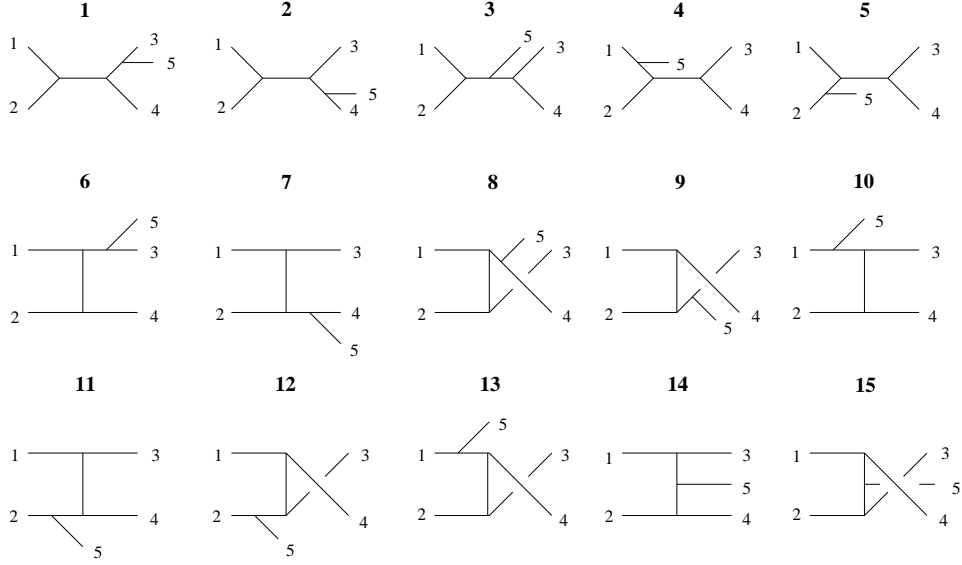}}
\caption{The fifteen three-vertex topologies contributing to the sum in the amplitude \eqref{eq:general_form}.
The labels $1$-$4$ correspond to the scalars and $5$ to the gluon.}
\label{fig:diagrams5}
\end{figure}

After resolving any quartic vertices into trivalent ones, the five-point gauge-theory amplitude can be written as a sum over 15 channels,
\begin{eqnarray}
\mathcal{A}_{5}=g^{3}\sum_{i=1}^{15}{c_{i}n_{i}\over d_{i}},
\label{eq:general_form}
\end{eqnarray}
where $c_{i}$ are the color factors defined by
\begin{eqnarray}
c_{1}={f}^{a_{5}a_{3}b}{f}^{ba_{4}c}{f}^{ca_{2}a_{1}}, & \hspace*{1cm} &
c_{2}={f}^{a_{5}a_{4}b}{f}^{ba_{3}c}{f}^{ca_{2}a_{1}}, \nonumber \\[0.2cm]
c_{3}={f}^{a_{2}a_{1}b}{f}^{ba_{5}c}{f}^{ca_{3}a_{4}}, & \hspace*{1cm} &
c_{4}={f}^{a_{5}a_{1}b}{f}^{ba_{2}c}{f}^{ca_{3}a_{4}}, \nonumber \\[0.2cm]
c_{5}={f}^{a_{5}a_{2}c}{f}^{ca_{1}b}{f}^{ba_{3}a_{4}}, & \hspace*{1cm} &
c_{6}={f}^{a_{5}a_{3}c}{f}^{ca_{1}b}{f}^{ba_{2}a_{4}}, \nonumber \\[0.2cm]
c_{7}={f}^{a_{5}a_{4}b}{f}^{ba_{2}c}{f}^{ca_{3}a_{1}}, & \hspace*{1cm} &
c_{8}={f}^{a_{5}a_{4}c}{f}^{ca_{1}b}{f}^{ba_{2}a_{3}}, \\[0.2cm]
c_{9}={f}^{a_{5}a_{3}b}{f}^{ba_{2}c}{f}^{ca_{4}a_{1}}, & \hspace*{1cm} &
c_{10}={f}^{a_{5}a_{1}b}{f}^{ba_{3}c}{f}^{ca_{2}a_{4}}, \nonumber \\[0.2cm]
c_{11}={f}^{a_{5}a_{2}b}{f}^{ba_{4}c}{f}^{ca_{3}a_{1}}, & \hspace*{1cm} &
c_{12}={f}^{a_{5}a_{2}b}{f}^{ba_{3}c}{f}^{ca_{4}a_{1}}, \nonumber \\[0.2cm]
c_{13}={f}^{a_{5}a_{1}b}{f}^{ba_{4}c}{f}^{ca_{2}a_{3}}, & \hspace*{1cm} &
c_{14}={f}^{a_{2}a_{4}b}{f}^{ba_{5}c}{f}^{ca_{3}a_{1}}, \nonumber \\[0.2cm]
c_{15}={f}^{a_{2}a_{3}b}{f}^{ba_{5}c}{f}^{ca_{4}a_{1}}, & \hspace*{1cm} &
\nonumber
\end{eqnarray}
with $f^{abc}$ being the structure constants. Finally, the denominators
\begin{eqnarray}
d_{i}=\prod_{\alpha_i}s_{\alpha_i},
\end{eqnarray}
correspond to the product of the kinematic invariants associated with the internal lines in the $i$-th diagram (the numbering
is the one shown in Fig. \ref{fig:diagrams5}).

Due to the Jacobi identities of the structure constants, the color factors satisfy nine independent identities that we label as $j_{\alpha}$. They are \begin{eqnarray}
j_{1}\equiv c_{12}-c_{9}+c_{15}=0, &\hspace*{0.5cm}& j_{2}\equiv c_{11}-c_{7}+c_{14}=0, \hspace*{1.cm}
j_{3}\equiv -c_{4}+c_{5}+c_{3}=0, \hspace*{0.5cm} \nonumber \\[0.2cm]
j_{4}\equiv c_{1}-c_{2}-c_{3}=0, &\hspace*{0.7cm}& j_{5}\equiv -c_{10}+c_{6}-c_{14}=0,\hspace*{0.7cm}
j_{6}\equiv -c_{13}+c_{8}-c_{15}=0, \label{eq:jacobi_identities}
\hspace*{0.6cm} \\[0.2cm]
j_{7}\equiv c_{4}-c_{10}+c_{13}=0, &\hspace*{0.5cm} & j_{8}\equiv c_{8}+c_{7}-c_{2}=0, \hspace*{1.3cm}
j_{9}\equiv c_{6}+c_{9}-c_{1}=0.
\nonumber  
\end{eqnarray}

The numerators $n_{i}$ in Eq.~\eqref{eq:general_form} can be computed from the Feynman rules of the theory. 
In general, they will not satisfy the Jacobi-like identities
$\pm n_{i}\pm n_{j}\pm n_{k}=0$, corresponding to $j_{\alpha}$ with $c_i\rightarrow n_i$.
However, this can be cured after performing a generalized gauge transformation, consisting of adding zero to the original amplitude in the form
\begin{eqnarray}
\mathcal{A}_{5}=\sum_{i=1}^{15}{c_{i}n_{i}\over d_{i}}+\sum_{\alpha=1}^{9}\gamma_{\alpha}j_{\alpha}
=\sum_{i=1}^{15}{c_{i}n'_{i}\over d_{i}},
\end{eqnarray}
where the new numerators $n'_{i}$ are obtained by collecting 
the coefficients of each color factor $c_{i}$ and multiplying by corresponding denominator: $n'_{i}=d_i \partial_{c_i}{\cal A}_5$. The parameters $\gamma_{\alpha}$ are unknown functions of the momenta and gluon polarization. 
They are determined by forcing the new numerators to satisfy the Jacobi identities
\begin{eqnarray}
j_\alpha\Big|_{c_i\rightarrow n'_i} =0.
\label{eq:jacobi_numerators}
\end{eqnarray}
These new numerators will be used to construct the corresponding gravitational amplitude using the BCJ double-copy prescription
\begin{eqnarray}
-i\mathcal{M}=\left({\kappa\over 2}\right)^{3}\sum_{i=1}^{15}{n_{i}'\widetilde{n}'_{i}\over d_{i}},
\label{eq:gravamp_BCJ}
\end{eqnarray}
with $\kappa$ being the gravitational coupling constant.

To study this amplitude in the multi-Regge kinematics limit it is convenient to express the momenta in terms of 
Sudakov parameters. As a first step we write 
\begin{eqnarray}
p_{3}=-p_{1}+k_{1}, \hspace*{1cm}  p_{4}=-p_{2}-k_{2}, \hspace*{1cm} p_{5}=-k_{1}+k_{2},
\end{eqnarray}
where in turn $k_{1}$ and $k_{2}$ are written as
\begin{eqnarray}
k_{1}^{\mu}=\alpha_{1}p_{1}^{\mu}+\beta_{1}p_{2}^{\mu}+k_{1,\perp}^{\mu}, \hspace*{1cm}
k_{2}^{\mu}=\alpha_{2}p_{1}^{\mu}+\beta_{2}p_{2}^{\mu}+k_{2,\perp}^{\mu},
\end{eqnarray}
with $k_{i,\perp}$ being vectors orthogonal to $p_1$ and $p_2$. Then, the gluon momentum takes the form
\begin{eqnarray}
p_{5}^{\mu}=(\alpha_{2}-\alpha_{1})p_{1}^{\mu}+(\beta_{2}-\beta_{1})p_{2}^{\mu}+k_{2,\perp}^{\mu}-k_{1,\perp}^{\mu}.
\end{eqnarray}
Finally, the multi-Regge kinematics regime is defined in terms of the Sudakov parameters by
\begin{eqnarray}
1\gg \alpha_{1} \gg \alpha_{2}, \hspace*{1cm} 1\gg |\beta_{2}|\gg |\beta_{1}|.
\label{eq:MRKlimit}
\end{eqnarray}
The gravitational amplitude \eqref{eq:gravamp_BCJ} can then be written as 
\begin{eqnarray}
-i\mathcal{M}=-iA_{kk}\mathcal{M}^{\mu\nu}\epsilon_{\mu\nu}(p_{5}),
\end{eqnarray}
where $\epsilon_{\mu\nu}(p_{5})$ is the graviton polarization tensor and \cite{SVSCVM}
\begin{eqnarray}
\mathcal{M}^{\mu\nu}&=&(k_{1}+k_{2})_{\perp}^{\mu}(k_{1}+k_{2})_{\perp}^{\nu}+
\mathcal{A}_{k1}\Big[(k_{1}+k_{2})_{\perp}^{\mu}p_{1}^{\nu}+p_{1}^{\mu}(k_{1}+k_{2})_{\perp}^{\nu}\Big] \nonumber \\
& & +\,\,\,\mathcal{A}_{k2}\Big[(k_{1}+k_{2})_{\perp}^{\mu}p_{2}^{\nu}+p_{2}^{\mu}(k_{1}+k_{2})_{\perp}^{\nu}\Big]
+\mathcal{A}_{12}\Big(p_{1}^{\mu}p_{2}^{\nu}+p_{2}^{\mu}p_{1}^{\nu}\Big) 
\label{eq:Mexpansion}
\\
& & +\,\,\,\mathcal{A}_{11}p_{1}^{\mu}p_{1}^{\nu}+\mathcal{A}_{22}p_{2}^{\mu}p_{2}^{\nu}.
\nonumber
\end{eqnarray}
The parameters $\mathcal{A}_{k1},\mathcal{A}_{k2},\mathcal{A}_{12},\mathcal{A}_{11},\mathcal{A}_{22}$ are to be determined. The coefficient $A_{kk}$ contains all the information about the coupling to external particles in the
amplitude\footnote{
Notice that this definition of $\mathcal{M}^{\mu\nu}$ differs from the one used in Ref. \cite{SVSCVM}
by the factorization of $A_{kk}$.}. By factoring it out we isolate Lipatov's graviton emission effective vertex 
$\mathcal{M}^{\mu\nu}$, which in multi-Regge kinematics 
is independent of the states involved in the collision and has the structure
\begin{eqnarray}
\mathcal{M}^{\mu\nu}=\Omega^{\mu}\Omega^{\nu}-\mathcal{N}^{\mu}\mathcal{N}^{\nu}.
\label{eq:lipatov_vertex}
\end{eqnarray}
The vectors $\Omega^{\mu}$ and $\mathcal{N}^{\mu}$ are given, respectively, by~\cite{Lipatov:1982vv,SVSCVM}
\begin{eqnarray}
\Omega^{\mu}\simeq \left(\alpha_{1}-{2\beta_{1}\over \beta_{2}}\right)p^{\mu}
+\left(\beta_{2}+{2\alpha_{2}\over \alpha_{1}}\right)q^{\mu}-(k_{1}+k_{2})^{\mu}_{\perp},
\end{eqnarray}
which corresponds to the effective vertex for the coupling of two reggeized gluons to an on-shell gluon, and
\begin{eqnarray}
\mathcal{N}^{\mu}\simeq -2i\sqrt{\beta_{1}\alpha_{2}}\left({p^{\mu}\over\beta_{2}}+{q^{\mu}\over \alpha_{1}}
\right).
\end{eqnarray} 
The term $\mathcal{N}^{\mu}\mathcal{N}^{\nu}$ is crucial for the cancellation of the simultaneous poles 
in $\alpha_{1}=0$ and $\beta_{2}=0$.

\paragraph{Distinguishable scalars.}

We first deal with the scattering of two distinguishable scalars $\Phi$ and $\Phi'$.
In Ref.~\cite{SVSCVM2} three of us analyzed this 
problem and found that the BCJ prescription only reproduces the QCD-like part of the gravitational amplitudes; that is, the $\Omega^{\mu}\Omega^{\nu}$ piece in Eq.~(\ref{eq:lipatov_vertex}).
In this paper we note that the problem with the incorrect $\mathcal{N}^{\mu}\mathcal{N}^{\nu}$ term is solved by embedding the Yang-Mills + 2 scalar theory into the bosonic sector of 
$\mathcal{N}=2$ super-Yang-Mills theory, which amounts to taking both scalars to transform in the adjoint representation and introducing
a matter self-coupling for the two scalars of the 
form 
\begin{eqnarray}
\Delta\mathcal{L}={g^{2}\over 2}{\rm Tr\,}\Big([\Phi,\Phi']^{2}\Big).
\label{eq:newcoupling_dist}
\end{eqnarray}
The corresponding Feynman rules then contain a new quartic scalar contact vertex and we need to add four more diagrams to the ones computed in Ref.~\cite{SVSCVM2}.
These new diagrams do not contain any $t$-channel poles, but their numerators combine with the remaining ones
to contribute to the double-copy amplitude in the Regge limit.
Evaluating all contributions, we find the following values for the numerators $n_{i}'$
\begin{eqnarray}
n_{1}'&=& (p_{1}+p_{2})^{2}\Big[-(\gamma_{9}-\gamma_{4})(p_{3}+p_{5})^{2}-2p_{3}\cdot\epsilon(p_{5})\Big],
\nonumber \\
n_{2}'&=& (p_{1}+p_{2})^{2}\Big[-(\gamma_{4}+\gamma_{8})(p_{4}+p_{5})^{2}+2p_{4}\cdot \epsilon(p_{5})\Big],
\nonumber \\
n_{3}'&=& (\gamma_{3}-\gamma_{4})(p_{1}+p_{2})^{2}(p_{3}+p_{4})^{2}, \nonumber  \\
n_{4}'&=& (p_{3}+p_{4})^{2}\Big[(\gamma_{7}-\gamma_{3})(p_{1}+p_{5})^{2}+2p_{1}\cdot \epsilon(p_{5})\Big], 
\nonumber \\
n_{5}'&=& -(p_{3}+p_{4})^{2}\Big[-\gamma_{3}(p_{2}+p_{5})^{2}+2p_{2}\cdot \epsilon(p_{5})\Big], \nonumber
\\
n_{6}'&=& -(p_{3}+p_{5})^{2}\Big[-(\gamma_{5}+\gamma_{9})(p_{2}+p_{4})^{2}+(p_{2}-p_{4})\cdot \epsilon(p_{5})\Big] 
\nonumber \\
& & -\,\,\,2(p_{2}-p_{4})\cdot (p_{1}-p_{3}-p_{5})[p_{3}\cdot\epsilon(p_{5})], 
\nonumber
\label{eq:numerators_disting}\\
n_{7}'&=& -(p_{4}+p_{5})^{2}\Big[(\gamma_{2}-\gamma_{8})(p_{1}+p_{3})^{2}+(p_{3}-p_{1})\cdot \epsilon(p_{5})\Big]
\nonumber \\
& &- \,\,\, 2(p_{3}-p_{1})\cdot (p_{2}-p_{4}-p_{5})[p_{3}\cdot\epsilon(p_{5})], 
\nonumber  \\
n_{8}'&=& (p_{2}+p_{3})^{2}\Big[(\gamma_{6}+\gamma_{8})(p_{4}+p_{5})^{2}+2p_{4}\cdot\epsilon(p_{5})\Big],
\label{eq:numdistprime} \\
n_{9}'&=& -(p_{1}+p_{4})^{2}\Big[(\gamma_{1}-\gamma_{9})(p_{3}+p_{5})^{2}+2p_{3}\cdot\epsilon(p_{5})\Big],
\nonumber \\
n_{10}'&=& -(p_{1}+p_{5})^{2}\Big[(\gamma_{5}+\gamma_{7})(p_{2}+p_{4})^{2}+(p_{2}-p_{4})\cdot\epsilon(p_{5})\Big]
\nonumber \\
& & -\,\,\, 2(p_{2}-p_{4})\cdot(-p_{1}+p_{3}-p_{5})[p_{1}\cdot\epsilon(p_{5})], \nonumber \\
n_{11}'&=& -(p_{2}+p_{5})^{2}\Big[-\gamma_{2}(p_{1}+p_{3})^{2}+(p_{3}-p_{1})\cdot\epsilon(p_{5})\Big] \nonumber \\
& & -\,\,\,2(p_{3}-p_{1})\cdot (-p_{2}+p_{4}-p_{5})[p_{2}\cdot\epsilon(p_{5})], \nonumber 
\\
n_{12}'&=& (p_{1}+p_{4})^{2}\Big[\gamma_{1}(p_{2}+p_{5})^{2}+2p_{2}\cdot\epsilon(p_{5})\Big], \nonumber \\
n_{13}'&=& -(p_{2}+p_{3})^{2}\Big[(\gamma_{6}-\gamma_{7})(p_{1}+p_{5})^{2}+2p_{1}\cdot\epsilon(p_{5})\Big], \nonumber 
\\
n_{14}'&=& -(p_{2}-p_{4})\cdot(p_{1}+p_{3}-p_{5})[(p_{3}-p_{1})\cdot \epsilon(p_{5})] \nonumber \\
& & -\,\,\,
(p_{3}-p_{1})\cdot (p_{2}-p_{4})[(-p_{1}+p_{2}-p_{3}+p_{4})\cdot\epsilon(p_{5})] \nonumber \\
& & +\,\,\, \gamma_{2}(p_{1}+p_{3})^{2}(p_{2}+p_{4})^{2}-\gamma_{5}(p_{1}+p_{3})^{2}(p_{2}+p_{4})^{2} ,\nonumber \\
n_{15}'&=& (\gamma_{1}-\gamma_{6})(p_{2}+p_{3})^{2}(p_{1}+p_{4})^{2}. \nonumber 
\end{eqnarray}

Although, in principle, we have nine equations for the nine functions $\gamma_{\alpha}$, momentum conservation 
makes some of the nine conditions in Eq.~\eqref{eq:jacobi_numerators} linearly dependent. One convenient 
way to implement momentum conservation is
to express our momenta using the Sudakov parameters introduced above. 
Doing so, we find that the equation system is described by a $9\times 9$ 
matrix that has rank 5 and the solution can be written in terms of 4 independent variables that we take
to be $\gamma_{1}$, $\gamma_{3}$, $\gamma_{6}$ and $\gamma_{7}$:
\begin{eqnarray}
\gamma_{2} &=& {(p_{2}+2p_{3}+p_{4})\cdot\epsilon(p_{5})\over s\beta_{1}}
-\gamma_{1}{1+\beta_{2}\over \beta_{1}}-\gamma_{3}{-1+\alpha_{1}-\alpha_{2}+\beta_{1}-\beta_{2}\over \beta_{1}},
\nonumber \\[0.2cm]
\gamma_{4}&=& {2(p_{3}+p_{4})\cdot\epsilon(p_{5})\over s}+\gamma_{3}(1-\alpha_{1}+\alpha_{2}-\beta_{1}+\beta_{2})+
\gamma_{7}(\beta_{1}-\beta_{2}),  \\[0.2cm]
\gamma_{5}&=& {(-p_{2}+p_{4})\cdot\epsilon(p_{5})\over s\alpha_{2}}-\gamma_{3}
{1-\alpha_{1}+\alpha_{2}-\beta_{1}+\beta_{2}\over \alpha_{2}}+\gamma_{6}{1-\alpha_{1}\over \alpha_{2}}
-\gamma_{7}{\beta_{1}-\beta_{2}\over \alpha_{2}}, \nonumber \\[0.2cm]
\gamma_{8} &=& {2(p_{2}+p_{3})\cdot\epsilon(p_{5})\over s(\alpha_{1}+\beta_{1})}-\gamma_{1}{1+\beta_{2}\over
\alpha_{1}+\beta_{1}}+\gamma_{6}{1-\alpha_{1}\over \alpha_{1}+\beta_{1}}-\gamma_{7}{\beta_{1}-\beta_{2}\over
\alpha_{1}+\beta_{1}}, \nonumber \\[0.2cm]
\gamma_{9}&=& -{2(p_{2}+p_{3})\cdot\epsilon(p_{5})\over s(\alpha_{2}+\beta_{2})}+
\gamma_{1}{1+\beta_{2}\over \alpha_{2}+\beta_{2}}-\gamma_{6}{1-\alpha_{1}\over \alpha_{2}+\beta_{2}}.
\nonumber
\end{eqnarray}

After applying the BCJ prescription~\eqref{eq:gravamp_BCJ}, the four independent $\gamma$'s cancel out of the gravitational amplitude, so
we set them to zero from now on. Plugging the five remaining $\gamma$'s back in the numerators  
of Eq.~\eqref{eq:numerators_disting}, we construct the gravitational amplitude from Eq.~\eqref{eq:gravamp_BCJ}.
In the multi-Regge kinematics limit, the coefficients in Eq.~\eqref{eq:Mexpansion} take the form
\begin{eqnarray}
\mathcal{A}_{11} &\simeq & \alpha_{1}^{2} -{4\alpha_{1}\beta_{1}\over \beta_{2}}
+{4\beta_{1}^{2}\over \beta_{2}^{2}}+{4\alpha_{2}\beta_{1}\over \beta_{2}^{2}}+\ldots\,,
\nonumber \\[0.2cm]
\mathcal{A}_{22} &\simeq & \beta_{2}^{2} +{4\alpha_{2}\beta_{1}\over \alpha_{1}}+{4\alpha_{2}\beta_{1}\over \alpha_{1}^{2}}
+{4\alpha_{2}^{2}\over \alpha_{1}^{2}}+\ldots\,,
\nonumber \\[0.2cm]
\mathcal{A}_{12} &\simeq & \alpha_{1}\beta_{2} -2\beta_{1}+2\alpha_{2}+\ldots\,,
\\[0.2cm]
\mathcal{A}_{k1} &\simeq & -\alpha_{1}+{2\beta_{1}\over \beta_{2}}+\ldots\,,
\nonumber \\[0.2cm]
\mathcal{A}_{k2} &\simeq & -\beta_{2}-{2\alpha_{2}\over \alpha_{1}}+\ldots\,,
\nonumber 
\end{eqnarray}
where the ellipsis denote subleading contributions in the multi-Regge limit.
The above coefficients correctly reproduce the full form of Lipatov's effective graviton emission vertex 
shown in Eq.~\eqref{eq:lipatov_vertex}, including the $-\mathcal{N}^{\mu}\mathcal{N}^{\nu}$ piece that was
not correctly retrieved in the analysis of Ref. \cite{SVSCVM2}.

We note that away from the Regge-limit the amplitude obtained from Eq.~\eqref{eq:gravamp_BCJ} is still a valid gravitational amplitude, however, it includes one additional contact term of the form $\sim \kappa^2 g^{\mu \nu}\Phi'{}^2 \partial_\mu \partial_\nu \Phi^2$. This term typically appears in the bosonic sector of supergravity theories, or in dimensional reductions of $D>4$ gravities. 
If desired, this contribution can be subtracted from the amplitude, thus obtaining a gravity amplitude with minimally coupled scalars.

\paragraph{Indistinguishable scalars.}
A second possibility to avoid 
the problem of Ref.~\cite{SVSCVM2} is to consider the theory of a single adjoint scalar minimally coupled to 
a nonabelian gauge field and compute the scattering amplitude of two indistinguishable scalars
with a gluon emission. 
Again, the number of Feynman diagrams contributing to the
amplitude is larger than the original calculation, since all channels are allowed. Resolving the diagrams
containing four-leg vertices in terms of trivalent ones, we arrive at the following form of the 
numerators in Eq.~\eqref{eq:general_form}:
\begin{eqnarray}
n_{1} &=& -
(p_{3}+p_{5})^{2}[(p_{2}-p_{1})\cdot\epsilon(p_{5})]
-2(p_{2}-p_{1})\cdot (-p_{3}+p_{4}-p_{5})[p_{3}\cdot\epsilon(p_{5})], \nonumber \\
n_{2}&=& -(p_{4}+p_{5})^{2}[(p_{2}-p_{1})\cdot\epsilon(p_{5})]
-2(p_{2}-p_{1})\cdot (p_{3}-p_{4}-p_{5})[p_{4}\cdot\epsilon(p_{5})], \nonumber \\
n_{3}&=& -(p_{2}-p_{1})\cdot(p_{3}-p_{4})[(p_{1}+p_{2}-p_{3}-p_{4})\cdot\epsilon(p_{5})] \nonumber \\
& & -\,\,\,(p_{3}-p_{4})\cdot(-p_{1}-p_{2}+p_{5})[(p_{2}-p_{1})\cdot\epsilon(p_{5})] \nonumber \\
& & -\,\,\,(p_{2}-p_{1})\cdot(p_{3}+p_{4}-p_{5})[(p_{3}-p_{4})\cdot\epsilon(p_{5})] ,\nonumber \\
n_{4}&=& -(p_{1}+p_{5})^{2}[(p_{3}-p_{4})\cdot \epsilon(p_{5})]-2(p_{3}-p_{4})\cdot(-p_{1}+p_{2}-p_{5})[p_{1}\cdot\epsilon(p_{5})], \nonumber \\
n_{5}&=& -(p_{2}+p_{5})^{2}[(p_{3}-p_{1})\cdot\epsilon(p_{5})]-2(p_{3}-p_{1})\cdot(-p_{2}+p_{4}-p_{5})
[p_{2}\cdot\epsilon(p_{5})], \nonumber \\
n_{6}&=& -(p_{3}+p_{5})^{2}[(p_{4}-p_{1})\cdot\epsilon(p_{5})]-2(p_{2}-p_{1})\cdot(-p_{3}+p_{4}-p_{5})
[p_{3}\cdot\epsilon(p_{5})], \nonumber  \\
n_{7}&=& -(p_{4}+p_{5})^{2}[(p_{3}-p_{1})\cdot \epsilon(p_{5})]-2(p_{3}-p_{1})\cdot(p_{2}-p_{4}-p_{5})
[p_{4}\cdot\epsilon(p_{5})],  \\
n_{8}&=& -(p_{4}+p_{5})^{2}[(p_{2}-p_{3})\cdot\epsilon(p_{5})]-2(p_{2}-p_{3})\cdot(p_{1}-p_{4}-p_{5})
[p_{4}\cdot \epsilon(p_{5})], \nonumber \\
n_{9}&=& -(p_{3}+p_{5})^{2}[(p_{4}-p_{1})\cdot\epsilon(p_{5})]-2(p_{4}-p_{1})\cdot(p_{2}-p_{3}-p_{5})
[p_{3}\cdot\epsilon(p_{5})], \nonumber \\
n_{10}&=& -(p_{1}+p_{5})^{2}[(p_{2}-p_{4})\cdot\epsilon(p_{5})]-2(p_{2}-p_{4})\cdot(-p_{1}+p_{3}-p_{5})
[p_{1}\cdot\epsilon(p_{5})], \nonumber \\
n_{11}&=& -(p_{2}+p_{5})^{2}[(p_{3}-p_{1})\cdot\epsilon(p_{5})]-2(p_{3}-p_{1})\cdot(-p_{2}+p_{4}-p_{5})
[p_{2}\cdot\epsilon(p_{5})], \nonumber \\
n_{12}&=& -(p_{2}+p_{5})^{2}[(p_{4}-p_{1})\cdot\epsilon(p_{5})]-2(p_{4}-p_{1})\cdot(-p_{2}+p_{3}-p_{5})
[p_{2}\cdot\epsilon(p_{5})], \nonumber \\
n_{13}&=& -(p_{1}+p_{5})^{2}[(p_{2}-p_{3})\cdot\epsilon(p_{5})]-2(p_{2}-p_{3})\cdot(-p_{1}+p_{4}-p_{5})
[p_{1}\cdot\epsilon(p_{5})], \nonumber \\
n_{14}&=& -(p_{2}-p_{4})\cdot(p_{1}+p_{3}-p_{5})[(p_{3}-p_{1})\cdot\epsilon(p_{5})] \nonumber \\
& & -\,\,\,(p_{3}-p_{1})\cdot(-p_{2}-p_{4}+p_{5})[(p_{2}-p_{4})\cdot\epsilon(p_{5})] \nonumber \\
& & -\,\,\,(p_{3}-p_{1})\cdot(p_{2}-p_{4})[(-p_{1}+p_{2}-p_{3}+p_{4})\cdot\epsilon(p_{5})], \nonumber\\
n_{15}&=& -(p_{2}-p_{3})\cdot(p_{4}-p_{1})[(-p_{1}+p_{2}+p_{3}-p_{4})\cdot\epsilon(p_{5})] \nonumber \\
& & -\,\,\,(p_{4}-p_{1})\cdot(-p_{2}-p_{3}+p_{5})[(p_{2}-p_{3})\cdot\epsilon(p_{5})] \nonumber \\
& & -\,\,\,(p_{2}-p_{3})\cdot(p_{1}+p_{4}-p_{5})[(p_{4}-p_{1})\cdot\epsilon(p_{5})].\nonumber
\end{eqnarray}

Remarkably, these numerators obtained from the application of the Feynman rules automatically satisfy the Jacobi-like identities~\eqref{eq:jacobi_numerators}, so
there is no need to perform a generalized gauge transformation.  We immediately proceed to construct the gravitational amplitude using the BCJ prescription \eqref{eq:gravamp_BCJ}. After taking
the multi-Regge kinematics limit \eqref{eq:MRKlimit} we find the following value for the coefficients in Eq.~\eqref{eq:Mexpansion}
\begin{eqnarray}
\mathcal{A}_{11} &\simeq & \alpha_{1}^{2} -{4\alpha_{1}\beta_{1}\over \beta_{2}}
+{4\beta_{1}^{2}\over \beta_{2}^{2}}+{4\alpha_{2}\beta_{1}\over \beta_{2}^{2}}+\ldots
\nonumber \\[0.2cm]
\mathcal{A}_{22} &\simeq & \beta_{2}^{2} +{4\alpha_{2}\beta_{1}\over \alpha_{1}}+{4\alpha_{2}\beta_{1}\over \alpha_{1}^{2}}
+{4\alpha_{2}^{2}\over \alpha_{1}^{2}}+\ldots
\nonumber \\[0.2cm]
\mathcal{A}_{12} &\simeq & \alpha_{1}\beta_{2} -2\beta_{1}+2\alpha_{2}+\ldots
\\[0.2cm]
\mathcal{A}_{k1} &\simeq & -\alpha_{1}+{2\beta_{1}\over \beta_{2}}+\ldots
\nonumber \\[0.2cm]
\mathcal{A}_{k2} &\simeq & -\beta_{2}-{2\alpha_{2}\over \alpha_{1}}+\ldots
\nonumber 
\end{eqnarray}
which again reproduce Lipatov's effective vertex \eqref{eq:lipatov_vertex}.

\section{Discussion}
 
In this note we have addressed the problem of applying color-kinematics duality to the 
scattering of two distinguishable scalar matter particles with gluon emission, or graviton emission. The calculation of Ref.~\cite{SVSCVM2}
suggested that, when applied to the scattering of minimally coupled distinguishable scalars, 
the BCJ double-copy prescription only reproduces
part of the gravitational amplitude in multi-Regge kinematics.
Here we have studied two extensions of the theory for which the prescription works, and which do not change the Regge limit studied in Ref.~\cite{SVSCVM2}. One consists of introducing a contact interaction between the two scalar particles, as suggested by the bosonic sector of $\mathcal{N}=2$ super-Yang-Mills theory. The second is to give up distinguishability of the scalars. In both cases the introduction of new diagrams contributing to the process gives valid gravity amplitudes from the BCJ double-copy prescription, and recovers the precise results of Ref.~\cite{SVSCVM2} in the Regge limit. 
  
The two cases can be thought of as originating from the bosonic sector of $D=4$ $\mathcal{N}=2$ super-Yang-Mills theory, keeping either both scalars, or only one scalar. Moreover, they can be regarded as coming from subsectors of $\mathcal{N}=4$ super-Yang-Mills theory, for which double-copy prescription is proven to give valid gravity tree-level amplitudes~\cite{BDHK}.
Since fermions do not play any role in the tree-level amplitudes studied here, the obtained results can be fully explained by supersymmetry. However, we note that supersymmetry is not a mandatory explanation of the results, nor is it the most elegant one. 

Dimensional reduction of Yang-Mills theories in $D>4$ provides a more direct path for understanding the results.
 Indeed, the interaction term \eqref{eq:newcoupling_dist} is generated by dimensionally reducing $D=6$ pure Yang-Mills to $D=4$, 
where the gauge field along the extra two dimensions are interpreted as two scalars,
$\Phi\equiv A_{4}$, $\Phi'\equiv A_{5}$.
The additional components of the gauge field strength tensor are
\begin{eqnarray}
F_{\mu 4}=D_{\mu}\Phi, \hspace*{1cm} F_{\mu 5}=D_{\mu}\Phi', \hspace*{1cm} F_{45}=-ig[\Phi,\Phi'],
\end{eqnarray}
with $D_{\mu}$ being the adjoint covariant derivative. The four-dimensional Lagrangian is then
\begin{eqnarray}
\mathcal{L}=-{1\over 4}{\rm Tr\,}\Big(F_{\mu\nu}F^{\mu\nu}\Big)+{1\over 2}{\rm Tr\,}\Big(D_{\mu}\Phi D^{\mu}\Phi\Big)
+{1\over 2}{\rm Tr\,}\Big(D_{\mu}\Phi' D^{\mu}\Phi'\Big)+{g^{2}\over 2}{\rm Tr\,}\Big([\Phi,\Phi']^{2}\Big).
\end{eqnarray}
Similarly, the tree-level scattering of two indistinguishable scalars with gluon emission can be computed either from this Lagrangian or from the dimensional reduction of
$D=5$ pure Yang-Mills to $D=4$ dimensions, which results in Yang-Mills theory with one adjoint scalar, where $\Phi\equiv A_{4}$. In the latter case there is no quartic scalar term.

Hence, one can say that the successful application of color-kinematics duality in the cases studied in our work directly stems from its validity in higher-dimensional Yang-Mills theory and gravity~\cite{BCJ,BCJLoop,BDHK,LoopNumerators,N=4SG}. 
Indeed, the dimensional reduction that we have used is a standard computational tool; it is frequently used in loop calculations in gauge and gravity theories with extended supersymmetry ({\it e.g.} see \cite{vm_SYM}).

While the modified theories we have considered avoid
the specific problem observed in Ref.~\cite{SVSCVM2}, 
the inclusion of general matter states and interactions in the color-kinematics formalism is still an open problem. Specifically, it would be interesting to understand how to precisely relate tree amplitudes in Yang-Mills theory with minimally-coupled matter, $N_s$ scalars and $N_f$ fermions, to that of Einstein gravity with similar matter content. We expect that an extension of the BCJ prescription is needed, which at intermediate steps embeds the gauge and gravity theories into their respective higher-dimensional versions. The results presented here, with the help of the information gained by taking the multi-Regge limit, were a first step towards understanding the general matter case.

\section*{Acknowledgments} 

A.S.V. acknowledges partial support from the European Comission under contract LHCPhenoNet (PITN-GA-2010-264564), 
the Madrid Regional Government through Proyecto HEPHACOS ESP-1473, the Spanish Government 
MICINN (FPA2010-17747) and Spanish MINECOs ``Centro de Excelencia Severo Ochoa" Programme under grant  SEV-2012-0249. 
The work of E.S.C. 
has been supported by a Spanish Government FPI Predoctoral Fellowship and grant FIS2012-30926. 
M.A.V.-M. acknowledges partial support from Spanish Government grants FPA2012-34456 and FIS2012-30926, 
Basque Government Grant IT-357-07 and Spanish Consolider-Ingenio 2010 Programme CPAN (CSD2007-00042).
M.A.V.-M. thanks the Instituto de F\'{\i}sica Te\'orica UAM/CSIC for kind hospitality.

\end{document}